\documentstyle[aps,preprint]{revtex}
\def\prl#1#2#3{{ Phys. Rev. Lett.} {\bf #1}, #2 (#3)}

\def\pre#1#2#3{Phys. Rev. E {\bf #1}, #2 (#3)}
\def\prb#1#2#3{Phys. Rev. B {\bf #1}, #2 (#3)}

\def\pr#1#2#3{Phys. Rev. {\bf #1}, #2 (#3)}
\def\prp#1#2#3{Phys. Rep. {\bf #1}, #2 (#3)}
\def\prlon#1#2#3{Proc. Phys. Soc. London A {\bf #1}, #2 (#3)} 
\def\rmp#1#2#3{Rev. of Mod. Phys. {\bf #1}, #2 (#3)}
\def\jpa#1#2#3{J. Phys. A {\bf #1}, #2 (#3)}
\def\jpc#1#2#3{J. Phys. C {\bf #1}, #2 (#3)}
\def\jpha#1#2#3{J. Phys. A {\bf #1}, #2 (#3)}

\def\physd#1#2#3{Physica D {\bf #1}, #2 (#3)}

\def\progm#1#2#3{Prog. Theor. Phy. Suppl. {\bf #1}, #2 (#3)}
\def\intjb#1#2#3{Int. J. Mod. Phys. B{\bf #1}, #2 (#3)}

\def\epc{$\epsilon_c$}
\def\eps{$\epsilon$}
\def\om{$\omega$}

\def\etl{$et ~al.$~}

\def\beqr{\begin{eqnarray}}
\def\eqnr{\end{eqnarray}}
\def\beq{\begin{equation}}
\def\bc{\begin{center}}
\def\ec{\end{center}}
\def\eqn{\end{equation}}
\begin{document}
\title{Dynamics of the Harper map: Localized states, Cantor spectra and
Strange nonchaotic attractors}
\author{Surendra Singh Negi and Ramakrishna Ramaswamy} 
\address{School of Physical Sciences\\ Jawaharlal Nehru
University, New Delhi 110 067, INDIA}
\date{\today}
\maketitle
\maketitle
\begin{abstract}
The Harper (or ``almost Mathieu'') equation plays an important role
in studies of localization.  Through a simple transformation, this
equation can be converted into an iterative two dimensional
skew--product mapping of the cylinder to itself.  Localized states of
the Harper system correspond to fractal attractors with nonpositive
maximal Lyapunov exponent in the dynamics of the associated Harper
map. We study this map and these strange nonchaotic attractors (SNAs)
in detail in this paper.  The spectral gaps of the Harper system have
a unique labeling through a topological invariant of orbits of the
Harper map. This labeling associates an integer index with each gap,
and the scaling properties of the width of the gaps as a function of
potential strength, $\epsilon$ depends on the index. SNAs occur in a
large region in parameter space: these regions have a tongue--like
shape and end on a Cantor set on the line $\epsilon = 1$ where the
states are critically localized, and the spectrum is singular
continuous. The SNAs of the Harper map are described in terms of
their fractal properties, and the scaling behaviour of their
power--spectra. These are created by unusual bifurcations and differ
in many respects from SNAs that have hitherto been studied. The
technique of studying a quantum eigenvalue problem in terms of
the dynamics of an associated mapping can be applied to a number of
related problems in 1~dimension.  We discuss generalizations of the
Harper potential as well as other quasiperiodic potentials in this
context.
\end{abstract}

\section{INTRODUCTION }

The Harper equation, the discrete Schr\"odinger equation for a 
particle in a quasiperiodic potential, 
\beq
\psi_{n+1} + \psi_{n-1} + 2\epsilon \cos 2\pi (n\omega+\phi_0) \psi_n
= E \psi_n, \label{harpq}
\eqn 
was originally introduced in the study of the motion 
of an electron in two dimensional lattice in a magnetic field 
\cite{Harper}.  In the above equation, $\psi_n$ is the wavefunction 
at lattice site $n$, $\omega$ is an irrational number, and 
$\phi_0$ is a parameter. The nature of the eigenfunctions and 
the eigenspectrum has been of considerable interest since 
this equation arises in numerous  and varied contexts 
\cite{soukl,wiegmann,hofstadter,Hilke,wig,mw}. 

Studies of localization phenomena have been important 
in condensed matter physics research for more than forty years 
\cite{anderson}, and here the Harper system has played a very
important role.  In systems with a periodic potential,
the Bloch theorem holds and leads to wavefunctions which typically
are extended. In the contrasting case of a random potential, the 
states are localized. A quasiperiodic potential is intermediate 
between these limits, and it is of interest to determine whether 
states are localized or extended. Indeed, the problem of localization 
has important implications in a variety of fields, among
the most significant being an explanation of transport properties in
disordered solids \cite{soukl,wiegmann,hofstadter,Hilke,wig,mw,anderson}
 (for example the Mott theory of the
metal--insulator transition \cite{mott-twose}). In a 1--dimensional 
tight-binding model with incommensurate modulation, at a critical 
strength of potential there exists a metal insulator transition.
Such a quasiperiodic system is physically realizable, for instance, 
in quasicrystals which have incommensurate modulation and therefore 
lack translational order \cite{Azbel}).

As is evident, the Harper equation is
invariant under the transformation $\psi_{n} \rightarrow (-1)^{n}
\psi_{n}, \epsilon \rightarrow - \epsilon, E \rightarrow -E$, so that
it suffices to consider the case $E, \epsilon$ nonnegative. A number
of studies \cite{wig,Azbel,simon,soko,jitomiriskaya,thou,geisel,Sun,aubry} 
have examined different properties of the eigenstates and spectrum of the
Harper equation. The most important feature of this system 
is that for  $\epsilon = 1$, the equation is self--dual \cite{aubry}: 
the equation for the Fourier transform of $\psi_n$ is identical to 
Eq.~(\ref{harpq}). The quantum eigenstates are extended for 
$\epsilon < 1$, exponentially localized for
$\epsilon > 1$ and but at the {\it critical} value $\epsilon = 1$,
the states are power--law localized.  The eigenvalue spectrum is
singular--continuous (namely, the spectrum forms a Cantor set) for
all irrational values of $\omega$ at this critical value of
$\epsilon$ \cite{hofstadter}. For rational values of $\omega$, the
Bloch theorem applies, and leads to a spectrum consisting of bands.

In a study of the continuous version of Eq.~(\ref{harpq}), namely
\begin{equation}
\label{qu}
-\frac{d^2\psi(x)}{dx^2}+2 \epsilon \cos(2\pi\omega x) \psi(x) ~~=~~
E \psi(x),
\eqn
Bondeson, Antonsen and Ott \cite{bao} pointed out that by using
Pr\"ufer's transformation, the above Schr\"odinger equation could be
converted into an ordinary differential equation for the 
variable $y = \tan^{-1} {(2g \psi^{\prime} \psi)/((\psi^{\prime})^2-
g^2\psi^2)}$, $g$ being an arbitrary constant, as
\beq
\label{cl}
\frac{dy}{dx}~~=~~\frac{1}{g}\{[g^2+k^2(x)]+[g^2-k^2(x)] \cos y \}
\eqn
where $k^2(x)$ = $E- 2\epsilon \cos (2\pi \omega x)$. This equation
describes a forced dynamical system (the independent
variable $x$ is now the ``time''), the forcing term deriving from
the potential being quasiperiodic if $\omega$ is irrational.
Study of the orbits of Eq.~(\ref{cl}) gives
information on the behaviour of wave--functions of Eq.~(\ref{qu}).
Indeed, the Lyapunov exponent $\lambda$ for an orbit of the dynamical
system is related to the localization length $\gamma$ of the
corresponding wavefunction \cite{bao},
\beq
\label{inv}
\lambda = -2/\gamma.  
\eqn
Thus localised states(finite $\gamma>$0) correspond to attractors 
with negative Lyapunov exponents, while extended states
($\gamma \to \infty$) correspond to orbits with zero
Lyapunov exponent. As Bondeson \etl \cite{bao} further argued, the
quasiperiodic driving precluded the possibility of a simple structure
for these attractors, which were both fractal and had nonpositive
Lyapunov exponents: these are strange nonchaotic attractors (SNAs)
\cite{gopy}.

Our concern in this paper is the dynamics of the Harper mapping which
is a similar transformation of Eq.~(\ref{harpq}), using Ricatti
variables, namely $\psi_{n-1}/\psi_{n} \to x_n$ \cite{luck,sk}.
Eq.~(\ref{harpq}) becomes the two--dimensional mapping
\begin{eqnarray}
\label{ks}
x_{k+1} &=& - [x_k - E + 2 \epsilon \cos 2 \pi \phi_k ]^{-1}\\
\phi_{k+1} &=& \phi_k + \omega \mbox{~~~~~~~mod~} 1. \label{ks2}
\end{eqnarray}
While this is exactly equivalent to the Harper equation, it is
amenable to somewhat simpler analysis since boundary conditions that
must be imposed for quantization in Eq.~(\ref{harpq}) become
conditions on the dynamical state of orbits in the mapping,
Eqs.(\ref{ks}-\ref{ks2}).

Viewed merely as an iterative mapping of the plane, the above system
has several features of interest. The map is reversible, so there can
be no chaotic dynamics, but for sufficiently large $\epsilon$, it can
be argued \cite{ks} that there must be an attractor of the dynamics
which has an infinite set of discontinuities. The attractors are
therefore both strange (i.e. fractal) and nonchaotic (the Lyapunov
exponents must be nonpositive): they are SNAs. The Harper map is
another example of a quasiperiodically forced nonlinear dynamical
system where it is known that SNAs are frequently encountered
\cite{rmp}. By now a variety of quasiperiodically driven systems with
SNA dynamics are known; these include examples of driven pendulums,
forced nonlinear maps, the driven van der Pol oscillator, neural
network systems, as well as some mappings where there is no explicit
quasiperiodic driving \cite{rmp}.  Owing to the fractal structure of
the attractors, the dynamics is strictly aperiodic, but the
correlation properties \cite{pf} of SNAs show them to be much less
aperiodic than chaotic attractors, the correlation function always
lying between 0 and 1.

Strange nonchaotic dynamics is of considerable interest since it is
intermediate in several ways between regular and chaotic motion.
SNAs are typically found in the region of a transition to chaos, and
several routes or scenarios for their formation are now known 
\cite{rmp}. Their existence can be confirmed by explicit calculation
of Lyapunov exponents and fractal dimensions, or by computing their
spectral and correlation properties.  In the Harper system it is
known that SNAs can be created by the homoclinic collision of an
invariant curve with itself \cite{prss}, and that this route to SNA
is accompanied by an unusual symmetry breaking. There are other
routes through which SNAs can also be formed, and some of these are
described here as well.

The Harper system is therefore of interest from both the nonlinear
dynamics perspective, as well as in the context of localization.
Indeed, as initial studies have shown \cite{ks,prss,nr}, both these
aspects of the problem are related, and the correspondence between
SNAs and localized states can be exploited so as to provide new
insights in the study of forced nonlinear systems, as well as
for localization studies.
It has been noted for instance, that quasiperiodic potentials are
frequently associated with critical or power--law localized states,
just as quasiperiodic driving frequently transforms chaotic
attractors to SNAs. This can be made more concrete by studying the
iterative mappings analogous to Eqs.~(\ref{ks}-\ref{ks2}) that are
associated with such potentials. In many instances, it also turns out
that such systems have attractors which are SNAs \cite{pramana}. 

This paper is concerned with such correspondences. One focus is on
what can be learned about the spectrum and eigenstates of the Harper
equation through a study of the dynamics of the Harper map. At the
duality point, $\epsilon = 1$, the eigenspectrum is given exactly by
the zeros of the Lyapunov exponent, $\lambda(E)$. Furthermore,
topological properties of the orbits of the mapping provide a
unique labeling of the gaps in the spectrum. The gap indices also
determine the scaling behaviour of the gaps as a function of the
coupling constant \cite{nr}.  These results are described in Section
III of this paper.  The behaviour of the gaps as a function of
coupling constant is important for a number of problems, and we 
have shown \cite{nr} how the gaps can be uniquely labeled by 
mapping them on to a Cayley
tree. The gap labels are winding numbers, which are topological
invariants for orbits of the iterative mapping. These determine the
scaling behaviour of the gaps at, above, and by duality, below, the
critical coupling, $\epsilon = \epsilon_c \equiv 1$. We discuss the
general scheme for labeling the gaps for particular choices of
$\omega$ such as the golden mean or silver mean ratios, as well as
for other algebraic and transcendental irrational ratios.

At the same time,
the SNAs of the Harper map are themselves of interest. Regions (in
parameter space) where they occur have an interesting hierarchical
geometry, being organized in fractal ``tongue--like'' zones
\cite{nr}. In Section II of this paper, we first describe properties
of the iterative mapping for the Harper potential. The phase diagram
for this system, as a function of $E$ and $\epsilon$ for fixed
irrational $\omega$ is discussed. The different bifurcation
routes for the formation of SNAs, and the measures have been
introduced in the past few years in order to distinguish SNAs from
other related attractors are elaborated in Section IV.

The case of quasiperiodic potentials obtained by generalizing the
Harper potential, where critical states are known to occur, are
discussed in Section V. We show that the technique of determining
the spectrum from the behaviour of the Lyapunov exponent versus
parameter curve can be extended to these potentials as well.  Other
cases of systems with critical states are known, as for example the
quasiperiodic potentials that arise from the Fibonacci, Thue--Morse,
Rudin--Shapiro and period--doubling symbolic sequences. 
The corresponding iterative mappings also support SNAs \cite{cal}. 
The paper concludes with a summary in Section~VI.

\section{The Harper map}

We study the mapping Eqs.~(\ref{ks}-\ref{ks2}) and other
quasiperiodic generalizations of it which arise from quasiperiodic
Schr\"odinger problems. While most of the essential features of the
system do not depend on the precise value of $\omega$ so long as it
is irrational, some of the details (such as the arrangements of the
gaps, see Section \ref{gaps} below) do. In this work $\omega$ is
taken to be a quadratic irrational number such as the golden mean ratio
$(\sqrt 5 -1)/2$, which is the value for which most previous studies
have been carried out, or the silver mean ratio, $\sqrt 2 -1$.

The Harper map is a skew--product dynamical system on the cylinder
{\bf S$^1 \otimes$ R}.  Since the $\phi$ dynamics is quasiperiodic,
there are no periodic orbits. There is also no chaotic dynamics. In
general there are three qualitatively different types of motion that
can occur in this system. There can be quasiperiodic tori
(two--frequency or three--frequency) with negative Lyapunov exponent,
neutrally stable curves with zero Lyapunov exponent, when the phase
space is conserved under the flow, and strange nonchaotic dynamics.
On SNAs the motion is both aperiodic and, as is characteristic of
such attractors, intermittent \cite{heagy}.

The $\phi$- freedom is neutral with the corresponding Lyapunov
exponent being trivially zero, while for the $x$- freedom, the
stretch exponent or one--step Lyapunov exponent \cite{pre} is
\beq
y_{i}=\ln |x^{\prime}_{i+1}| = 2 \ln |x_{i+1}|.
\eqn	
The $N$-step finite--time Lyapunov exponent is
\beq
\label{loclyap}
\lambda_N (x_i) = {1 \over N} \sum_{j=0}^{N-1} y_{j+i}.
\eqn
These local exponents are indexed by initial condition $x_{i}$, but
the asymptotic Lyapunov exponent,
\beq
\label{lyap}
\lambda = \lim_{N \to \infty}\lambda_N (x_i),
\eqn
is independent of (almost every) initial condition. In the rest of
the paper, we deal only with this nontrivial Lyapunov exponent, for
which the following relation holds \cite{simon,aubry},
\beq
\lambda(E,\epsilon) =-2 \ln (\epsilon) + \lambda(1/\epsilon,E/\epsilon).
\eqn
relating the exponents above and below the critical coupling.

The eigenvalue $E$ appears simply as another parameter in the Harper
map.  Quantization conditions that apply for the Harper equation,
namely that the wavefunctions should be normalizable, now become
conditions on the dynamical state of the map. In particular, these
can be viewed as conditions on the Lyapunov exponents: for extended
states the Lyapunov exponent is zero, while for localized states, the
Lyapunov exponent is negative.

At the duality point \epc, eigenstates of the Harper equation are
critically localized. Hence, the Lyapunov exponent of the Harper map
is identically zero whenever $E$ is an eigenvalue. Furthermore, since
the spectrum is known to be singular--continuous, the $\lambda$
versus $E$ curve meets the line $\lambda = 0$ on a Cantor set of
points (Figs.~1a,b).  Indeed, the Hofstadter butterfly
\cite{hofstadter} which is the
spectrum of the Harper equation as a function of $\omega$ can be
determined as the set $\{ E: \lambda(E,\omega,\epsilon_c) = 0\}$.

The variation of the Lyapunov exponent with $\epsilon$ is simple if
$E$ is an eigenvalue. Below $\epsilon = 1$, the exponent is zero and
the state is extended, while above it, $\lambda \sim - \ln \epsilon$
and the state is localized; see Fig.~2 for the case of $E=0$. Of
particular note is the discontinuity (in slope) at \epc: the
localization transition can therefore be viewed as a bifurcation. As
shown previously, this bifurcation is accompanied by a quasiperiodic
symmetry--breaking and the homoclinic collision of neutrally stable
invariant curves to form a SNA
\cite{prss}. For other eigenvalues, the location of the eigenvalue
changes with $\epsilon$. At \epc~there is always a
symmetry--breaking, but there need not be the homoclinic collision,
so SNAs can be formed here by other routes \cite{negi,kk} as is discussed
in Section \ref{scenarios} below.

The overall dynamical behaviour in the Harper map can be summarized
in a phase diagram, shown in Fig.~3a for the case of $\omega = 
\sqrt{2} -1$, namely the silver mean ratio. Owing to the
$E \to -E$ symmetry, it suffices to consider $(E > 0, \epsilon > 0)$.
We compute the Lyapunov exponent and determine the dynamical state in
the different regions demarcated in the figures. Below \epc, the
dynamics is either on invariant curves (I) or on tori (Q), while
above \epc, there are regions of SNA (S) and tori (Q). The SNA
regions exist only at and above \epc. As was first argued by Ketoja
and Satija \cite{ks}, in the strong--coupling limit of large \eps,
the dynamics is guaranteed to be on SNAs; this argument can be
extended down to \epc. Above \epc, therefore, the dynamics is largely
on SNAs. The gap regions extend above \epc. Therefore, by continuity,
the SNA regions emanate from the eigenvalues at \epc. These have a
tongue--like shape, and since each eigenvalue is associated with one
such region, these are hierarchically organized in the same manner as
are the energy eigenvalues: SNAs occur in fractal tongues. Shown in
Fig.~3b is a (schematic) detail in the neighborhood of one (and therefore of 
every) tongue: the region of 3--frequency quasiperiodic orbits or 
tori for $\epsilon < 1$ are mirrored by SNA regions for $\epsilon > 1$.

The major differences between the cases of different $\omega$ arise
from the manner in which the regions of tori (Q) are arranged.
These correspond to gaps in the eigenspectrum of the Harper operator,
which we now discuss.

\section{Gaps of the Harper spectrum}
\label{gaps}
The eigenvalue spectrum of the Harper equation is
singular--continuous for irrational \om~at \epc, namely it forms a
Cantor set. The nature of the gaps in the spectrum has been of great
interest in a number of different contexts in the past several years.
In particular, how the gaps behave as a function of the potential
strength, namely \eps, has been a long--standing question \cite{luck,thouless}.

The eigenvalues can be simply deduced from the $\lambda$ versus $E$
curve at \epc, as shown in Fig.~1a for the case of $\omega$ the
golden mean ratio, or in Fig.~1b for $\omega$ the silver mean ratio.
The tori regions of the Harper map correspond to gaps in the
eigenvalue spectrum of the Harper equation. Orbits for such values
of $E$ and \eps~ are 1--d curves that wind across the cylinder an
integral number of times. Within each gap, this integer index is
constant and measures the number of times the orbit crosses the
boundaries $x \to \pm \infty$. This essentially counts the number of
changes of sign of the wave--function (per unit length), and is
related to a winding number for the orbit. These indices are
indicated in Figs.~1a and 1b for the largest visible gaps.

For rational frequencies, the spectrum of the Harper equation consists 
of a finite number of bands. If $\omega = p/q, p,q $ integers, then there
are $q$ or $q-1$ bands depending on whether $q$ is odd or even, and therefore
in the positive energy spectrum the number of gaps is $(q-1)/2$ or $(q/2-1)$.
For rational $\omega$ all orbits are periodic with period $q$. For
energies in the gaps, the orbits form  a 1--parameter family, indexed
by initial phase, from which the winding number can be easily extracted.

As is well--known \cite{numb}, an irrational number has a unique continued
fraction representation. Thus an irrational frequency 
\beq
\omega \equiv [a_1,a_2, a_3, \ldots] = 
{1\over{a_1+ {1\over{a_2 + {1\over{a_3+ {1\over \ddots}}}}}}}
\eqn
$a_i$'s integers, 
can be successively approximated by a unique series of rational 
approximants. By truncating the continued fraction representation
up to the term $a_j$, one obtains the rational $p_j/q_j$; the 
sequence $p_j/q_j, j=1,2, \ldots$ converges, 
$ \lim_{k\to\infty} p_k/q_k \to \omega $. For a given 
approximant $p_j/q_j$, the gap structure of the spectrum can 
be described in terms of the winding numbers. These run from
1 to $q_j/2 -1$ or $(q_j-1)/2$, as discussed  in the preceding 
paragraph. The new gaps that appear for the next approximant, 
$p_{j+1}/q_{j+1}$, do not alter the {\it inter se} ordering of the
gaps already present for lower order rational approximants. Thus
the gap structure for irrational $\omega$ can be deduced by
sequentially examining the structure of the gaps for the
successive rational approximations to $\omega$.
Each gap in the spectrum, or each region of tori in the map, can
therefore be uniquely labeled by an integer index. 

For algebraic irrationals which have periodic continued
fraction representations, there are simplifications that
permit a more compact description of the ordering scheme.
We now describe this for the case for the golden and silver
mean ratios.

The golden mean is the limiting ratio of successive Fibonacci
numbers, $F_k/F_{k+1}$, where $F_k$ is given by the relation
\beq
F_{k+1} = F_k + F_{k-1}, k= 1,2,\ldots, \mbox{~with~} F_0 = 0, F_1=1,
\eqn
the first few being 0,1,1,2,3,5,8,\ldots, while the silver mean is
the limiting ratio of the successive integers of the family
0,1,2,5,12,29,70,\ldots, namely
\beq
S_{k+1}= 2S_k + S_{k-1}, k= 1,2,\ldots, \mbox{~with~} S_0 = 0, S_1=1
\eqn
with $\lim_{k\to\infty} S_k/S_{k+1}\to \sqrt 2 -1$.

The ordering of the gaps can be best described by first constructing
a graph for the particular irrational ratio in the following manner.
We discuss this in detail for the silver mean ratio below, and
indicate the analogous result for the golden mean ratio \cite{nr}.

Consider a Cayley tree, arranged as shown in Fig.~4a, with each node
(except the origin, labeled 0) having two successors. Nodes at the
same horizontal level are at the same generation.  The rightmost node
at each generation is labeled by successive $S_k$, while the leftmost
are half the successive even $S_k$'s. The other integers are
identified with the remaining nodes as follows: for given $m$, the
mother node $i_m$ is the smallest available such that
\begin{itemize} 
\item 
$m+i_m$ is one of the integers $S_k$. If $i_m$ already has two
daughters, then the mother node is chosen such that
\item
$m+i_m$ is one of the integers $\bar S_{k+1} = S_{k+2} - S_{k+1}$,
namely 1,3,7,17,41,\ldots.
\end{itemize}
For $\omega$ the golden mean, $i_m$ is chosen such that $m+i_m$ is
the smallest possible Fibonacci integer \cite{nr}.  Complicated (but
deducible \cite{nr}) rules determine which branch (left or right) the
daughters are placed, but the above procedure results in an unique
arrangement of all integers on the Cayley tree: see Figs.~4a and 4b.

The actual ordering of the gaps derives from the following additional
consideration. Every pair of integers, $i_1$ and $i_2$, with $i_2 >
i_1$, has two possibilities as to how they are relatively placed on
this graph. Either
\begin{enumerate}
\item
$i_1$ is an ancestor of $i_2$, i.e. there is a directed path
connecting $i_2$ to $i_1$. If this path is to the left at node $i_1$,
then $i_2 \prec i_1$. (If to the right, then $i_1 \prec i_2$.)\\
\noindent or
\item
$i_0$ is the most recent common ancestor of $i_1$ and $i_2$. If the
path from $i_0$ to $i_1$ is on the left at $i_0$, then $i_1 \prec
i_2$. (Similarly, if it is to the right, then $i_2 \prec i_1$.)
\end{enumerate}
This gives a unique ordering of the integers with the relation
$\prec$ being transitive (if $i \prec j$ and $j \prec m$ then $i\prec
m$). Thus, for the silver mean we get $$\ldots \prec 35 \prec \ldots
\prec 11 \prec
\ldots \prec 1 
\prec \ldots \prec 4 \prec \ldots \prec 2 \prec \ldots \prec 
7 \prec \ldots \prec 12 \prec \ldots \prec 0.$$ while for the golden
mean, the ordering is $$\ldots \prec 4 \prec \ldots \prec 9 \prec
\ldots \prec 1 
\prec \ldots \prec 7 \prec \ldots \prec 2 \prec \ldots \prec 
11 \prec \ldots \prec 21 \prec \ldots \prec 0.$$

The gaps appear in precisely this order: if $k \prec
\ell$, then the gap with index $k$ precedes the gap with index $\ell$
in the positive energy spectrum of the Harper system: see Figs.~1a,b
where gap labels for the largest visible gaps are indicated. The
labels have been determined by examination of orbits of the map as
described. The attractor shown in Fig.~5 is an orbit in the gap labeled 8 (which
lies between gaps 4 and 3) in the silver--mean Harper system.

As can be seen in Figs.~1a and ~1b, each gap has two attributes: the
width and the depth. The width is the spacing between two energy
eigenvalues of the Harper equation, but the depth has no immediate
significance, being just the minimum value that the Lyapunov exponent
takes between two zeros. These are denoted $w_m$ and $d_m$
respectively, and both are functions of the coupling
strength, \eps.

In general, both widths and depths decrease with order (i.e. with
increasing gap index), and with increasing \eps. However, and
particularly in the case of the widths, the dependence on gap label
is nonmonotonic, in a manner described below. The depths scale in a
simple fashion, namely
\beq
\label{depth}
d_N \sim {C \over {N \epsilon^N}},
\eqn
as shown in Figs.~6a,b. On the other hand, the widths must be considered in 
{\it families}
labeled along a consistent branch of the Cayley tree. For example,
for $\omega = \sqrt 2 -1$ (see Fig.~4a), the gaps 1,2,5,12,29,\ldots,
the rightmost nodes at each generation form one family, while gaps
1,6,35,204,\ldots, the leftmost nodes, form another family, for each
of which the widths scale, at \epc, as
\beq
\label{width}
w_N \sim {1 \over {N^{\theta}}},
\eqn
the exponent $\theta$ being particular to a given family of gaps. The
two families of gaps enumerated above are the fastest and slowest
decreasing, respectively, with exponents $\theta_l \approx 2.05$ and
$\theta_r \approx 2.31$. For all other families, the exponents lie
within the limits [$\theta_l,\theta_r$]. Similar results for the
golden mean case are shown in Fig.~5b.

Wiegmann \etl \cite{wig} have studied the tight--binding Harper
Hamiltonian for which they obtain the gap distribution, namely the
number of gaps lying between $D$ and $D+dD$. They find that this
quantity, $\rho(D) \sim D^{-\gamma}$ where $\gamma$ is found to be
approximately $3/2$; similar exponents have been reported by Geisel
\etl
\cite{geisel} as well.
If the gaps are rank--ordered, namely from largest downwards,
disregarding the gap indices, then we observe that they decrease as
\beq
\label{rankwidth}
w_r \sim {1 \over {r^\mu}}.
\eqn
$\mu \approx 2$, which is consistent with the power--laws obtained
earlier
\cite{wig,geisel}. 

\section{SNAs in the Harper system}
\label{scenarios}
We now turn to the description of the dynamical attractors in the Harper map.
Strange nonchaotic dynamics in the Harper system is known to be
created through a number of different routes or scenarios. Unlike
other examples of quasiperiodically forced systems wherein SNAs are
formed by mechanisms that have parallels with routes to chaos in
similar unforced systems, here the routes to SNA can be quite
distinctive.

\subsection{Homoclinic collision and symmetry breaking}

As has been described earlier, at band center, namely for $E=0$,
orbits of the Harper map below \epc~lie on invariant curves that
consist of two branches (these are denoted C for central and N for
noncentral in Fig.~7a). These branches originate in the fact that
the mapping for $\epsilon = 0$ has a period--2 fixed point and no
period--1 fixed point. Different initial points on the $(x,\phi)$
plane generate different invariant curves, the two branches of which
are widely separated at \eps=0, but as $\epsilon \to 1$, these
branches begin to approach each other.  For a particular curve, the
branches collide at a dense set of points, at \epc, and indeed, all
the different curves merge at \epc~ to give a single attractor which
has a dense set of singularities (Fig.~7b). At the merging point, the
Lyapunov exponent remains zero. The distance between the two branches
of a given invariant curve decreases as a power in \epc-\eps, as does
the distance between any pair of invariant curves (Fig.~7c).
As previously noted, the homoclinic collision route to SNA
is accompanied by a symmetry--breaking. Below \epc~when the Lyapunov
exponent is zero, there is a quasiperiodic symmetry in the stretch
exponents (Fig.~7d), and this symmetry is lost above \epc, once the
attractor has negative Lyapunov exponent (Fig.~7e).

While each eigenvalue at \epc~is associated with a SNA, not all SNAs
arise from the above homoclinic collision route. Indeed, the more
general route to SNA in this system appears to be related to both the
blowout bifurcation as well as the so--called fractalization scenario
for the formation of SNAs, and this is discussed in the following
subsection.

\subsection{Other Bifurcations and Fractalization}

Given the complexity of the phase diagram for the Harper system with
the hierarchical organization of SNA (S) and torus (Q,G) regions, there are
a number of bifurcations as parameters are varied. For fixed
$\epsilon$ below \epc, there are bifurcations from 2--frequency to
3--frequency tori as $E$ is varied, the signature of this transition
being the change of the Lyapunov exponent from a negative value to
zero. There is therefore a discontinuity in the slope of the Lyapunov
exponent versus $E$ curve.  The 3--frequency tori (Q) densely cover the
phase space, while the 2--frequency tori (G) are attractors of the
dynamics (including the case when the Lyapunov exponent is exactly
zero). This transition is also, therefore, accompanied by a 
breaking of the symmetry in the stretch exponents.
On the other hand, by fixing $E$ and varying $\epsilon$, there are
other bifurcations depending on the value of $E$. There can be
transitions from 3--frequency tori to 2--frequency tori below
\epc, and from 2--frequency tori to SNA above \epc. 

Shown in Fig.~8a is the variation of the Lyapunov exponent with 
\eps~for $E=1$; the bifurcations being signaled by a
discontinuity in slope of the curve, with the Lyapunov exponent being
strictly nonpositive throughout. SNAs are first formed at this energy
at \eps $\approx$ 2.5200, below which there is an attracting
2--frequency torus. This transition has some of the dynamical
characteristics of the blowout bifurcation \cite{blowout} both in the
nature of the attractors before (Fig.~8c) and after (Fig.~8d) the
transition as well as in the manner in which the Lyapunov exponent
varies. The fluctuations in the local Lyapunov exponents, as 
measured in the variance of the distribution, are large on the
SNAs and zero on the tori. This is shown in Fig.~8b. Note that
the Cantor set structure of the gaps (whose widths decay only as
a power in \eps) is reflected in the pattern of 
bifurcations along any typical line in the ($E$,\eps) plane.

The G$\to$ S  transition occurs along essentially all the boundaries of the
SNA regions above \epc. When approached from below, the attractor of
the dynamics can be seen to gradually develop from a smooth curve to
one with wrinkles on all scales: this is reminiscent of the
fractalization process \cite{kk}. 
The blowout nature of the transition--insofar as the attractor
abruptly changes its volume with intermittent ``bursts''--can
be seen more clearly along the line $E = 3$.  
In the torus region G, the dynamical map here has a single
invariant curve (Fig.~9a), which develops kinks and wrinkles as \eps~is varied.
The transition to SNA is at $\epsilon \approx 1.270$(37567), when the
attractor gets wrinkled enough and actually hits the boundary $x \to
\infty$, Fig.~9b. This bifurcation can be clearly detected by measuring, say,
the distance $\Delta = \min_n \vert 1-\tanh x_n \vert$, namely the closest 
point of the attractor from the boundary as a function of \eps. As can 
be seen in Fig.~9c, this quantity abruptly decreases at the transition.

\subsection{Characterization of Harper SNAs: Correlations}

Principal among the measures that can be used to distinguish SNAs
from morphologically similar chaotic attractors or from (regular)
attractors which are similar dynamically (namely they have similar
values of the Lyapunov exponent) are quantities that examine
correlations on the attractors \cite{pf,jpa}.
                     
The autocorrelation function, which provides a quantitative measure
of the extent to which dynamical properties are correlated, is
defined for a dynamical variable $x$ as
\beq
C(\tau) = \frac{\langle x_i x_{i+\tau}\rangle
 -\langle x_i\rangle \langle x_{i+\tau}\rangle}
{\langle x_i^2\rangle -\langle x_i\rangle ^2} 
\eqn
where $i = 1,2,\ldots$ is a discrete time index, $\tau = 0,1,\ldots$
is the time shift, and $\langle \rangle$ denotes a time--average. If
$x$ is periodic then $C(\tau)$ varies between -1 and 1, regaining its
initial value $C(0)$ in a periodic fashion.
If $x$ is chaotic, then $C(\tau)$ decays from its
initial value of 1 to 0, around which latter value it fluctuates.
On SNAs, the variables have correlation functions
that do not recur exactly: $C(\tau)$ is also quasiperiodic and is nearly
recurrent, oscillating between -1 and 1 (Fig.~10a).  

The Fourier transform of the correlation function also shows these
differences as characteristic spectra. The Fourier transform of a
discrete sequence $\{x_k \}$ is
\beq
T(\Omega, N) =\frac{1}{N} \sum_{k=1}^N x_k \exp (i2\pi k \Omega),
\eqn
the power spectrum being
\beq
P (\omega) = \lim_{ N\to \infty} \langle \vert T(\omega, N) \vert^{2}
\rangle.
\eqn
A chaotic signal typically has a continuous power spectrum, but for
SNAs, the Fourier spectrum reflects
the fact that the dynamics is neither chaotic nor regular, and the
power spectrum consists of several peaks at frequencies corresponding to 
the quasiperiodicity in the dynamics; see Fig.~10b.

\subsection{Phase Sensitivity Properties}

Pikovsky and Feudel \cite{pf} introduced the phase--sensitivity
exponent in order to characterize the strangeness of a attractor.
This quantity measures the sensitivity of the dynamics with respect
to change of phase of the external force.  Given a map 
\begin{eqnarray}
x_{n+1}&=& f(x_{n},\theta_{n})\nonumber\\
\theta_{n+1} &=& \theta_{n}+\omega \mbox{~~~mod~} 1,
\end{eqnarray}
the maximal value of the derivative ${\partial x}/{\partial \theta
}\;$ can provide a suitable tool to distinguish between strange and
nonstrange geometry. Since the above derivative changes along the
trajectory $(x_0,\theta _0), (x_1, \theta _1), (x_2, \theta_2),
\ldots$,
one can get a relation
\begin{eqnarray}
S_{N} &=&\frac{\partial x_{n+1}}{\partial \theta}\nonumber\\
&=&f(x_{n},\theta_{n})+f(x_{n},\theta_{n}) \frac{\partial
x_{n}}{\partial
\theta }
\end{eqnarray}
where $S_N$ is the derivative with respect to external phase 
(we follow the notation of Pikovsky and Feudel \cite{pf}). The number 
of maxima in $S_N$ up to time $N$ is
\beq
\label{part}
\gamma_{N}(x,\theta )=\max_{0\leq n\leq N}\vert S_n \vert.
\eqn
The value of $\gamma _N$ grows with $N$, a consequence of the fact
that the attractor is not smooth; see Fig 11a.  Also, as
Pikovsky and Feudel \cite{pf} noted, the growth rate of the partial
sum with time represent the strangeness of attractor, so it becomes
necessary to calculate the minimum value of $\gamma_{N}(x,\theta )$
with respect to randomly choose initial points,
\beq
\Gamma_{N}= \min_{x,\theta}[~~ \max_{0\leq n\leq N}\vert S_n \vert ~~].
\eqn
From Fig.~11b we see that $\Gamma_{N}\sim N^{\mu}$, where $\mu$ is
the so--called phase sensitivity exponent. If the attractor is smooth
(i.e. nonfractal) then the maximum derivative with respect to 
external phase is bounded, and this value saturates as a function of
iteration.  However, in the case of SNAs, the maximum derivative with 
respect to external phase increases indefinitely with iteration.

\section{Maps of Harper type}

In this section we discuss the dynamics of mappings similar to 
Eqs.~(\ref{ks}-\ref{ks2}).
The Harper map can be generalized in two essentially different ways.
By generalizing the potential in the Harper equation but retaining
the feature of quasiperiodicity, one essentially obtains other
Schr\"odinger equations. Power--law localization is known to occur in
a variety of quasiperiodic systems, well studied examples being the 
potentials \cite{kohmoto}
\beq
V_n = \epsilon_1 \cos 2\pi n \omega + \epsilon_3 \cos 6\pi n \omega
\label{third}
\eqn
and
\beq
\label{tanh}
V_n = {\epsilon \over \tanh \mu} \tanh (\mu \cos 2\pi n \omega).
\eqn
In both these systems, it is known that there are mobility edges. 
In the resulting
maps, which are very similar to the Harper, the cosine ``driving''
term is replaced by a more general function which is also
quasiperiodic since $\omega$ is irrational. 

Since both these above cases can be viewed as modifications of the
Harper system, the phase diagrams obtained by variation of $E$ and
the parameters ($\epsilon_1, \epsilon_3$ or $\epsilon, \mu$) are
quite similar to Fig.~3a,b. Some instances have been studied
previously. For Eq.~(\ref{third}) with $\epsilon_3 = 1/4$, the
regions of 3-- and 2--frequency tori are arranged in a manner very
similar to the case with $\epsilon_3 = 0$, namely the Harper map
(Fig.~3a), except that the transition from extended to localized
states does not occur on the line $\epsilon_1 = 1$, but instead on
some complicated curve in the ($E, \epsilon_1)$ plane; see Fig.~12a.
Fig.~12b shows the corresponding phase diagram for the potential in 
Eq.~(\ref{tanh}).

A number of quasiperiodic potentials generated from abstract
sequences support critical states \cite{luck}. Among the most
extensively studied such sequences is the Fibonacci chain, the
potential for which is defined by
\beqr
V_n &=& \alpha~~~~~~~~~~~~~ 0 \le \{n\omega\} \le \omega \nonumber \\
&=& -\alpha~~~~~~~~~~~\omega < \{n\omega\} \le 1.
\label{fibo}
\eqnr
For any value of $\alpha$, all states of this system are critical
\cite{kohmoto,kohmoto1,kohmoto2,pandit}.  The corresponding
Schr\"odinger equation can be transformed to the mapping
\beq
x_{k+1} = - [x_k - E + V_k ]^{-1}
\eqn
with $V_k$ given by Eq.~(\ref{fibo}). Analogous maps can be obtained
for the Thue--Morse, Rudin--Shapiro and period--doubling
quasiperiodic sequences, and these have been shown to support SNAs
\cite{pramana}. The nature of the eigenspectrum in such cases will be
discussed elsewhere \cite{cal}. 

Another possible generalization of the Harper map gives 
\begin{eqnarray}
\label{ksg}
x_{k+1} &=& - [f(x_k) + 2 \epsilon \cos 2 \pi \phi_k ]^{-1}\\
\phi_{k+1} &=& \phi_k + \omega \mbox{~~~~~~~mod~} 1. \label{ksg2}
\end{eqnarray}
where $f(x)$ is now arbitrary while the quasiperiodic driving term is retained.
Because of the somewhat unusual nature of the mapping,
not much can be said in general---there can now be chaotic motion as
well if $f$ is not 1--1, for example. However, analysis of the
strong--coupling limit still suggests that SNAs should arise in such
systems. Indeed, for special cases that have been examined so far
\cite{prss,negi} many of the scenarios to SNA that have been seen in
the Harper mapping also exist in these more general cases. An
important limitation is that except for the case when $f$ is linear,
the above mapping cannot arise from a Schr\"odinger equation, and
thus the parallel insights that derive from study of the 
quantum-mechanical system are not available for nonlinear $f(x)$. 

\section{Summary and Discussion}
\label{conclu}

In the present work we have studied the Harper equation from a 
dynamical systems point of view. The correspondence between 
localized states and strange nonchaotic attractors 
\cite{bao,ks} makes it possible to obtain a number of properties 
of the eigenspectrum, as well as the eigenstates of the Harper 
equation through a study of the equivalent iterative Harper map.

For the spectral gaps, we have described a labeling scheme which 
explicitly connects the number--theoretic properties of $\omega$
with the gap ordering. This is most conveniently depicted on a 
Cayley tree, which we describe in detail for the case of the so--called
silver--mean quadratic irrational $\omega = \sqrt 2 -1$. The
general principles governing this organization are not difficult to 
ascertain, and computationally, these are most easily deduced through
examination of the orbits of the Harper map. The gap indices that
we obtain are related to the winding numbers introduced earlier by
Johnson and Moser \cite{jm}, and these (integer) indices determine the
scaling behaviour of the gaps as a function of the coupling. We find
that the width of each gap decreases only as a power of the coupling, and 
thus the gaps never close: this confirms the conjecture of Thouless \etl 
\cite{thouless}. The organization of the gaps on the Cayley tree also
makes it possible to identify other scaling properties. For fixed $\epsilon$,
the widths of gaps along a branch of the tree also show a power--law 
scaling in the gap indices.

The transition from extended to localized states is accompanied by a 
breaking of the symmetry of {\it stretch exponents} \cite{prss}.  This
symmetry--breaking appears to accompany a number of the bifurcations
in this system. Recall that the stretch exponents are the logarithms 
of the ratio of wavefunctions at adjacent sites. Thus the symmetry 
that gives a zero value of the Lyapunov exponent in the extended state 
is the following: the ratio of the wavefunctions at a triple of sites, 
$i-1, i, i+1$, (a:b:c, say) is such that at the triple of sites $j-1, j, j+1$,
the ratio of wavefunctions is (approximately, because of the quasiperiodicity) 
reversed (namely c:b:a), for each $j$ such that  $j-i$ is a Fibonacci number. 
On the other hand, the wavefunctions of the localized phase do not have 
this symmetry and are instead characterized through their multifractal 
oscillations, which have also been extensively studied \cite{multi}.

There have been several realizations of the Harper system in experiments.
Experiments  of wave--chaos in microwave scattering \cite{stockmann}  in
a geometry that is designed to mimic the Harper equation have shown the 
Hofstadter butterfly. In a very recent experimental study of the Hall 
conductance of a suitably designed heterostructure \cite{klitz}, the 
Cantor--set nature of the energy spectrum has been explicitly detected. 
The scaling results for the gaps reported in this paper and elucidation
of the gap organization are relevant to such studies.

The methods of analysis employed here have wide applicability. A number of
quasiperiodic systems in 1--dimension are known to have singular continuous 
spectra. The correspondences studied here for the Harper system appear to 
be valid in the general cases as well, and thus offer an alternate and 
powerful means of analyzing such problems.

\section*{Acknowledgment} This work is supported by a grant from the
Department of Science and Technology. We thank Prof. Jean Bellisard
for very helpful correspondence on the spectral gaps of the Harper
equation.

\newpage
\centerline{Figure Captions}
\begin{itemize}
\item[Fig. 1.]
(a) Lyapunov exponent versus energy at \epc~for $\omega = (\sqrt 5
-1)/2. $ Gap labels are indicated for the largest visible gaps.  
At every bifurcation, when $\lambda = 0$, the dynamics is on a SNA.  
(b) Lyapunov exponent versus energy at \epc~for $\omega =\sqrt 2 -1.  
$ Gap labels are indicated, as in Fig.~1a, for the largest visible gaps.
\item[Fig. 2.]
Lyapunov exponent versus $\epsilon$ for the eigenvalue $E=0$. The
extended to localized transition occurs at \eps=\epc=1.

\item[Fig. 3.]
(a) Phase diagram for the Harper map for $\omega = \sqrt 2 -1$.
Below \epc, namely the vertical line the dynamics is on
three--frequency quasiperiodic (Q) orbits or extended states, 1--d
attractors which are two--frequency quasiperiodic orbits or gaps (G),
and above \epc, the dynamics is on SNAs (S), and 1--d attractors (G).
Only the largest gaps are visible at this scale. 

(b) Detail in the neighborhood of the central gaps. Every tongue
of SNAs necessarily has a similar structure.

\item[Fig. 4.]
(a) Ordering of the gaps for $\omega = (\sqrt 5 -1)/2 $, the golden
mean ratio. Only part of the
Cayley tree described in the text is shown for clarity. Each node has
two daughters except for 0, which has only one.

(b) The corresponding Cayley tree for the case of $\omega$ the silver
mean, $\sqrt 2 -1$.

\item[Fig. 5.]
The attractor for a value of $E$ corresponding to the gap $N=8$.
Note that the orbit has 8 branches that traverse the range $-\infty <
x < \infty$.

\item[Fig. 6.]
Scaling of the gap widths, $w_N$ ($\bullet$), and depths $d_N$
($\diamond$) as a function of gap index, $N$, at \eps~= \epc~for the
cases of (a) $\omega = (\sqrt 5 -1 )/2$ and (b) $\omega =\sqrt 2 -1$.  
For clarity, 
the depths have been multiplied by a factor of 10.  
The dashed line fitting the depths has slope -1. 
The dotted lines show the scaling of the two families of
gaps; see the text for details.

\item[Fig. 7.]
(a) Invariant curve for the Harper map corresponding to the band
center extended state, the central and noncentral branches being
indicated by C and N respectively.  (b) The fractal attractor
corresponding to the critically localized state.  (c)
Scaling of the minimum vertical distance $d$ between a family of
different invariant curves in the Harper map as a function of the
parameter $(1 - \epsilon)$ for the eigenvalue $E = 0$. Different
pairs of curves approach each other at different rates, but at \epc,
{\it all} curves collide to form the critical SNA. (d) First return
map for the stretch exponent showing the quasiperiodic  symmetry
below \epc, and (e) the return map above \epc, when this symmetry is
broken. 

\item[Fig. 8.] The depressed blowout bifurcation along the line $E=
1$.  (a) Variation of the Lyapunov exponent with $\epsilon$ for
$\omega$ the golden mean. (b) Fluctuations of the Lyapunov exponent.
The attractor (c) before (a two--frequency torus) and (d) after (a
SNA) the bifurcation. 

\item[Fig. 9.]
The bifurcations along the line $E=3$, showing the attractor (a)
before ($\epsilon = 1.270375\ldots $) and (b) after the G $\to $ S transition. 
(c) The distance from the attractor to the boundary as a function of \eps.

\item[Fig. 10.] 

For the $E=0$ critical SNA for $\omega$ the golden mean (a) the
correlation function, and (b) the power spectrum.

\item[Fig. 11.] 

(a) Maxima of the partial sum $\gamma_{N}$ (cf. Eq.~(\ref{part})) on
the critical SNA at \epc, $E$=0 in the Harper map, as a function of
iteration length, $N$.  (b) $\Gamma_{N}$, the minimum value of
$\gamma_{N}$.

\item[Fig. 12.] 
Phase diagram for (a) the generalized Harper model, Eq.~(\ref{third})
with $\epsilon_1 = $ and $\epsilon_3$. The tongue--like regions
where SNAs can be found continue to be hierarchically arranged 
but the transition from localized
to extended states occurs along a curve in the plane  rather than
on the line \eps=1.
 (b) As in Fig.~12(a) for the potential of Eq.~(\ref{tanh}).

\end{itemize}

\end{document}